\documentclass[twocolumn,amssymb,amsmath,amsfonts,showpacs,prb,floatfix]{revtex4}

\usepackage{dcolumn}%
\usepackage{bm}%
\usepackage{graphicx}%

\def\dH{\Delta H(M,\Delta M)} 
\def\Neff{N_{\rm eff}} 

\def\Hr{H_{\rm r}} 
\def\Ha{H_{\rm a}} 
\def\Hi{H_{\rm i}} 
\def\Hc{H_{\rm c}} 
\def\Hb{H_{\rm b}} 
\def\Heff{H_{\rm eff}} 
\def\Hk{H_{\rm K}} 
\def\Hs{H_{\rm s}} 

\def\ha{h_{\rm a}} 
\def\heff{h_{\rm eff}} 
\def\hex{h_{\rm ex}} 

\def\Ku{K_{\rm u}} 
\def\Ms{M_{\rm sat}} 
\def\m0{\mu_0} %
\def\M2s{\Ms^2} 

\def\Masc{M^-} 
\def\Mdes{M^+} 
\def\Mrev{M^{\lrcorner}} 

\def\Dsi{D (H_{\rm s})} 

\def\DseFORC{D_{\rm FORC}(\Hc)} 

\def\sest{\sigma^{\ast}} 

\def\komm#1{}           

\begin{document}
\preprint{APL}

\title{Extracting the intrinsic switching field distribution in perpendicular media: a comparative analysis}

\author{Michael Winklhofer}%
\surname{Winklhofer} \email{MICHAELW@LMU.DE}
 \altaffiliation[Also at~]{Department of Earth and Environmental Science, Ludwig-Maximilians-Universit\"at, Theresienstr. 41, D-80333 M\"unchen, Germany}
\homepage{http://www.geophysik.uni-muenchen.de/michael}
\author{Gergely T. Zimanyi}%
\surname{Zimanyi}
\affiliation{Department of Physics and Astronomy, UC Davis, One Shields Avenue, Davis, CA 95616, USA}

\begin{abstract}
We introduce a new method based on the first-order-reversal-curve
(FORC) diagram to extract the intrinsic (microscopic)
switching-field distribution (SFD) of perpendicular recording
media (PRM). To demonstrate the viability of the method, we
micromagnetically simulated FORCs for PRM with known SFD and
compare the extracted SFD with the SFD obtained by means of two
different methods that are based on recoil loops, too, which
however rely on mean-field approximations and assumptions on the
shape of the SFD. The FORC method turns out to be the most
accurate algorithm over a broad range of dipolar interaction
strengths, where the other methods overestimate the width of the
SFD. \komm{The best match between the extracted and intrinsic
$\Dsi$ can be obtained in the case of $\pm45\deg$-tilted PM. Here,
there is some power in the 45 degrees lines from the center of the
distribution towards the reversible axis, but that can easily be
corrected for, for example, by means of SORCs. Also ( dM(Hr,
Ha)/Hr) gives an indication where the distribution 'really' is).
Here, meanfield correction works because the curvilinearity of the
loops is conserved. In the perpendicular plane, however, we cannot
undo the curvilinear hysterons into square hysterons ... well,
unless we do a vector Preisach model, maybe }

\end{abstract}



\maketitle

\section{INTRODUCTION}

The quality of recording media depends crucially on the intrinsic
(microscopic) switching-field distribution (SFD) of the media
particles, which determines both magnetic stability and attainable
recording density. It is straightforward to obtain the SFD of a
diluted magnetic system by taking the derivative of the DC
demagnetization (DCD) curve. In the case of high-density magnetic
recording media, where magnetic interactions between the media
particles are not negligible, the problem of extracting the SFD
from bulk magnetization curves cannot be solved rigorously any
more and the shape (and to a lesser extent the location) of the
extracted SFD will depend on certain model assumptions.

Two conceptually different algorithms \cite{Veerdonk:03,Berger:05}
have been suggested recently to extract the intrinsic $\Dsi$ of
perpendicular recording media (PRM) from macroscopic magnetization
curves using a set of recoil loops. The analysis technique by
\citet{Veerdonk:03} assumes a constant effective demagnetization
factor $\Neff$ for deshearing recoil loops, from which the DCD
curve is extracted. Extraction of the DCD curve and deshearing the
recoil loops are performed simultaneously in order to arrive at a
self-consistent solution\cite{Veerdonk:03}. Although the algorithm
converges after a couple of iterations, it is not clear how
reliable the self-consistent solution may be under the assumption
of an effective demagnetization factor independent of the
magnetization. Another drawback of the method is that it requires
assumptions about the shape of the SFD, which of course is not
known.

The second algorithm, referred to as $\dH$-method
\cite{Berger:05}, is a generalization of the $\Delta H$ method
originally proposed by \citet{TagawaNakamura:91}. The $\dH$-method
overcomes the restriction of a constant value of $\Neff$, assuming
an effective field of the form $\Heff(\Ha,M)=\Ha+\Hi(M)$, where
$\Ha$ and $\Hi$ denote the applied and internal field,
respectively, and $M$ is the magnetization in the direction of
$\Ha$. The $\dH$-method approximates interactions on the
mean-field level in a sense that all microscopic magnetization
configurations $\mathbf{M}(\mathbf{r})$ representing the same
macroscopic magnetization value $M$ produce the same
volume-averaged internal field $\Hi(M)$. An implicit assumption
underlying the $\dH$-method is that each particle acts a square
hysteron. Then, the lower branch of the major loop $M^-(\Ha)$ can
be represented as:
\begin{equation}
\Masc(\Ha) = - \Ms + 2 \int_{-\infty}^{\Heff(\Ha,M)} \Dsi d\Hs \,,
\label{Masc}
\end{equation}
where $\Dsi$ is the SFD. Any recoil loop $\Mrev(\Hr, \Ha>\Hr)$
originating from the upper branch $\Mdes$ at $\Hr$ can be written
as
\begin{equation}
\Mrev(\Hr, \Ha) = \Mdes(\Hr) + 2 \int_{-\infty}^{\Heff(\Ha,M) }
\Dsi d\Hs \,. \label{Mrecoil}
\end{equation}
In order for Eq.~\ref{Mrecoil} to hold, two conditions must be
met: firstly, $\Mrev(\Hr,\Ha>\Hr)$ has to saturate at fields
$|\Ha|\leq|\Hr|$ and secondly, recoil loops must not cross each
other. Then the coercivity distribution can be assumed to consist
of disjunct segments $\Dsi d\Hs$ (non-interacting hysterons). The
inverse $I^{-1}$ of the cumulative distribution
\begin{equation}
I (H) = \int_{-\infty}^{H } \Dsi d\Hs \, \label{cumdist}
\end{equation}
for a given magnetization value $M$ can then be obtained by taking
the difference in field position, $\Delta H (M)$, between
$M^-(\Ha)$ and recoil loop $\Mrev(\Hr, \Ha>\Hr)$,
\begin{eqnarray}
\dH \equiv \Ha(M^-) &-&\Ha(\Mrev)  = \label{dHeq} \\
= I^{-1}\left[\frac{\Ms + M }{2}\right]
 &-& I^{-1}\left[\frac{\Ms + (M -\Delta M) }{2}\right]
 \,, \nonumber
\end{eqnarray}
where $\Delta M=\Mdes(\Hr) - (-\Ms) $. By fitting the $\dH$ curves
against the inverse of a certain parameterized distribution
function, the key features of the SFD can be extracted
\cite{Berger:05}. In case $I$ is a normal distribution centered at
$\Hc$, Eq.~\ref{dHeq} writes to
\begin{equation}
\frac{\dH}{\Hc} = \sqrt{2}\sigma\left[{\rm erf}^{-1}(m) - {\rm
erf}^{-1}(m-\Delta m) \right] \label{GaussfitdH} \,,
\end{equation}
where lower case $m$'s are the magnetization values relative to
the saturation magnetization $\Ms$.

The third method in our comparative analysis is based on the
first-order-reversal curve (FORC)
distribution\cite{Mayergoyz:85,Pike:99}, defined as
\begin{equation}
\rho(\Hr,\Ha) = - \frac{1}{2} \frac{\partial^2
M(\Hr,\Ha)}{\partial \Hr \partial \Ha} \,. \label{defFORC}
\end{equation}
Although there is no principal difference between a FORC and a
recoil loop, the numerical determination\cite{Pike:99} of $\rho$
from a set of FORCs requires recoil loops measured on a grid
equidistant in $\Hr$ and $\Ha$.

For an assemblage of square hysterons, the FORC distribution is
identical to their Preisach distribution and we can obtain their
SFD as
\begin{equation}
\DseFORC = \int_{-\infty}^{\infty} \rho(\Hc,\Hb) d\Hb \,,
\label{DHsFORC}
\end{equation}
where $\Hc=(\Ha-\Hr)/2$ and $\Hb=(\Ha+\Hr)/2$ are the coordinates
of the Preisach plane, defined by the microscopic switching field
$\Hc$ and the bias field $\Hb$. As with the $\dH$ method, the FORC
method is reliable as long as the media particles can be
reasonably well described by square hysterons. As opposed to the
two algorithms above, however, the method based on
Eq.~\ref{DHsFORC} relies neither on a mean-field approximation nor
on a certain model form for $\Dsi$ and therefore imposes the
fewest constraints on the data.

\section{COMPUTATIONAL DETAILS}

The three methods presented above are best tested on a system with
known SFD. For this purpose, we micromagnetically computed sets of
FORCs for a PRM, using the OOMMF code\cite{OOMMF} (v. 1.1b2). The
simulated medium typically consisted of $N\sim2\cdot 10^3$
particles, arranged on a regular grid (mesh size $s=5$~nm),
inscribed in a circle of 250~nm diameter. We chose a circular
boundary to minimize the effects of corners. The easy axes of the
particles are all oriented the same way, roughly perpendicular to
the surface ($\theta_K=89.42\deg$) so as to avoid numerical
problems that may arise when the applied magnetic field $H$
($\theta_H=90\deg$) is exactly collinear with the easy direction.
We modified the module {\tt maganis.cc} to produce normal
distributed values of the uniaxial magnetocrystalline anisotropy
constant, $\Ku$ such that $\sigma_{\Hk}=0.1~\langle\Hk\rangle$,
with $\Hk=2~\Ku/\mu_0\Ms$. It has been shown
\cite{ZhouBertramSchabes:02} that a small amount of intergranular
exchange helps to reduce the increase in the recording transition
parameter due to a small distribution in $\Hk$. We therefore set
the nondimensional intergranular exchange coupling constant
$\hex=A/(\Ku\,s^2)$ to $\hex=0.046$ in all our simulations. In
order to systematically explore the effects of dipolar
interactions, we vary the dipolar interaction strength by using a
scaling factor $\phi<1$ for $\Ms$, where $\phi$ represents the
volume fraction of magnetic material in each cell. The uniaxial
magnetocrystalline anisotropy constant, $\Ku$, is scaled
simultaneously to keep the value of the microscopic coercivity
$\Hk$ constant as $\phi$ varies.\komm{A less efficient way of
decreasing the interaction strength in FFT-based codes would be to
increase the spacing between adjacent cells by introducing
nonmagnetic cells in between. Such an
approach
not only inflates the computation time and memory, but also limits
the range of possibles value of $\phi$.}
We normalize all magnetic fields by $\Hk$ and use a mean-field
type equation to define the material-independent dipolar
interaction strength as $\phi_Q =\phi/Q$, i.e.,
\begin{equation}
\heff = \ha + \alpha \frac{\phi}{Q}\frac{M}{\Ms} \equiv \ha +
\alpha\, \phi_Q m \,, \label{mfscaled}
\end{equation}
where $\alpha$ is the mean-field parameter and $Q$ denotes the
quality factor $Q=2\Ku/(\mu_0\M2s)$. For hcp Cobalt at room
temperature, $Q=0.422$. 
\komm{This choice of parameters and anisotropy direction yields
$694$~mT as a mean value of the intrinsic switching field
(compared to $\m0\Hk=743$~mT for the exactly collinear case).}

\section{RESULTS AND DISCUSSION}

\begin{figure}
\includegraphics[width=8.5cm]{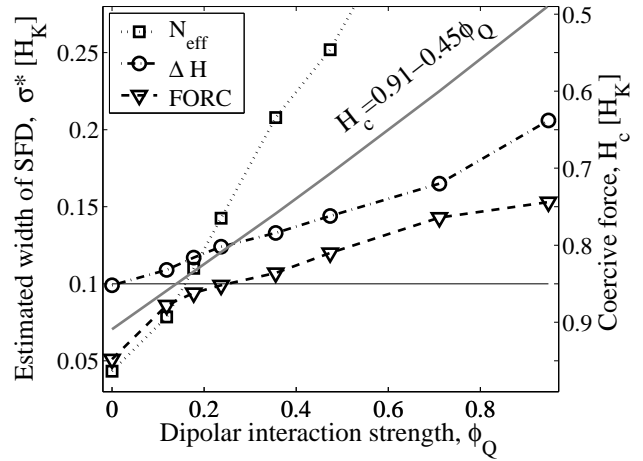}
\caption{\label{compareEstSigma}
Comparison of the three methods in terms of the extracted $\sest$
of the SFD as a function of the dipolar interaction strength
$\phi_Q$. The relative strength of the exchange coupling decreases
from left to right ($\hex=0.046$ for all values of $\phi_Q$). The
intrinsic SFD is a Gaussian distribution with $\sigma=0.1$ (thin
horizontal line). The coercive force is represented by the grey
line without plot symbols ($\Hc$ scale on the right).\komm{For
$\phi_Q>0.1$, the $\Neff$ method\cite{Veerdonk:03} yielded a
better fit when a lognormal distribution was assumed and therefore
the $\sigma$ corresponding to the best fitting lognormal
distribution is plotted here. Symbols: squares $\Neff$ method;
circles $\dH$ method; triangles FORC.} }\end{figure}

\begin{figure}
\includegraphics[width=8.5cm]{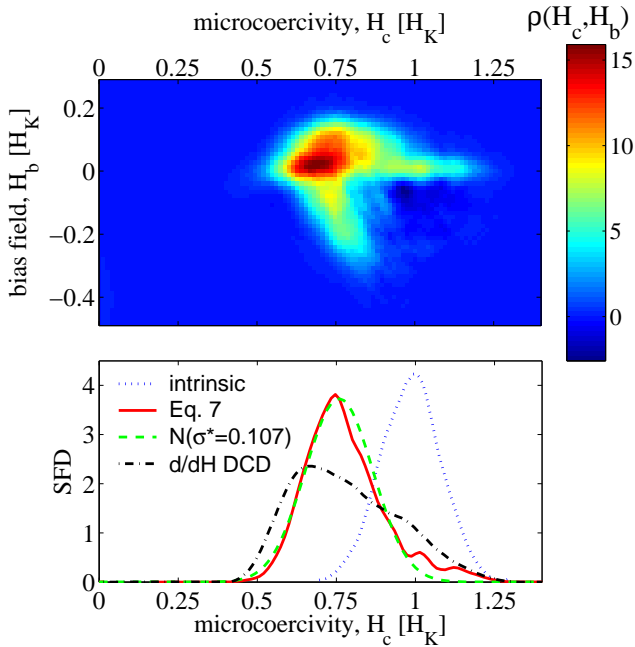}\\[1mm]
\caption{\label{NeffvsFORC} (Color online). FORC diagram
$\rho(\Hc,\Hb)$ (above) obtained for a PRM with $\phi_Q=0.35,
\hex=0.046$ and the corresponding SFD extracted on the basis of
Eq.~\ref{DHsFORC} (below). The width $\sest=0.107$ is determined
from fitting the extracted SFD (solid line) with a normal
distribution centered at $\Hc=0.76\,\Hk$ (dashed). The derivative
of the DCD curve (dash-dotted) has a distinct asymmetry. The
intrinsic SFD (dotted) with $\sigma=0.1$ is shown for comparison.
All curves are normalized to unit area. With the $\dH$ method,
$\sest$ was obtained as 0.133 (original recoil loops and $\dH$
curves available online).}\end{figure}

Figure~\ref{compareEstSigma} shows the estimated width $\sest$ of
the SFD as extracted with the three algorithms tested here. In the
hypothetical limit case of a PRM controlled solely by
exchange-coupling ($\phi_Q\leq 0$), the $\dH$ method is able to
capture the intrinsic width of the SFD with great accuracy. For
PRM largely dominated by dipolar interactions, however, the $\dH$
method is consistently less accurate than the FORC method, which
renders the least deviation of $\sest$ from $\sigma$ for $\phi_Q
> 0.15$. In contrast, the $\sest$ extracted with the $\Neff$ method always
show the largest deviations from $\sigma$. Importantly, the
$\Neff$ method does not yield self-consistent solutions any more
for $\phi_Q>0.25$, where the asymmetry of the DCD curve is too
pronounced to be properly described by a normal distribution
(Fig.~\ref{NeffvsFORC}). More consistent solutions in this regime
can be found with a log-normal SFD cut-off at high fields. Caution
should therefore be taken to not interpret asymmetric DCD curves
prematurely as evidence of an asymmetric intrinsic SFD. The FORC
method on the other hand captures the intrinsic shape of the SFD
to a very good degree despite the shift in location
(Fig.~\ref{NeffvsFORC}).

All the three methods are primarily concerned with finding the
right scale parameter (i.e., $\sest$ or FWHM) of the SFD
distribution, while the location parameter is considered to be
invariant. As can be seen in Figure~\ref{compareEstSigma}, the
observed values of $\Hc$ are progressively shifted to lower values
with increasing $\phi_Q$, which is due to dipolar interactions
between the media particles, deflecting the effective field away
from the easy axes. According to the Stoner-Wohlfarth
relationship,
\begin{equation}
\Hs (\psi)  = \Hk \, \left((\cos\psi)^{2/3}
+(\sin\psi)^{2/3}\right)^{-3/2}
\end{equation}
observed coercive forces of $\sim 0.8 \Hk$ suggest that a large
fraction of grains experience a local effective field that
deviates by some 5\% from the applied field direction, giving rise
to curvilinear hysterons, in other words, reversible magnetization
processes. In the FORC diagram (Fig.~\ref{NeffvsFORC}),
curvilinear hysterons manifest themselves in the form of the small
negative region centered at $(\Hc,\Hb)=(1,-0.05)\Hk$. This way,
one can determine directly from the FORC diagram if and to what
degree the model assumptions are not strictly met.

Since all methods tested here rely on square hysterons, their
underlying assumption starts breaking down with increasing dipolar
coupling $\phi_Q$, albeit at different values of $\phi_Q$. More
importantly, the fact that the methods do not all start to fail at
the same point also shows that the presence of reversible
magnetization processes is not the most crucial limitation to a
method's applicability, which rather is restricted by imposing
constraints on the internal field distribution. Compared to the
$\Neff$ method, the $\dH$ method puts a less tight constraint on
the expected magnetization curves and so has a larger range of
applicability. The very absence of any such approximations makes
the method based on the FORC distribution the most robust
algorithm and therefore most suitable for characterizing
ultra-high-density PRM.

%
We enjoyed discussion with Chris Pike, Kai Liu and Rene van der
Veerdonk. This work was funded by the Campus-Laboratory Exchange
Program (UC).

\newpage

\section*{Online Supplements}

\vbox{\includegraphics[width=8.5cm]{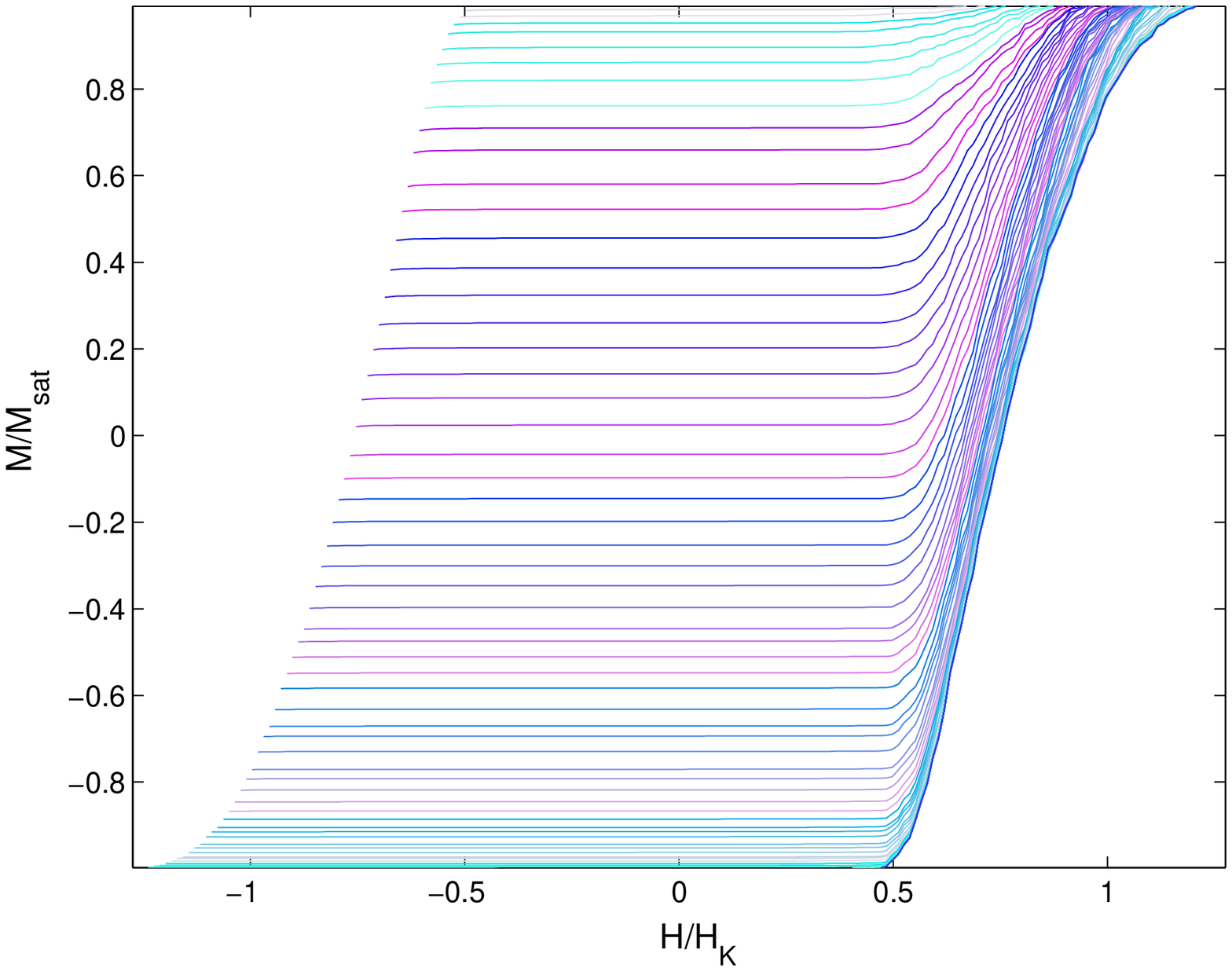}\\[1mm]
Micromagnetically computed FORCs for $\phi_Q=0.35$.}

\vspace{0.4cm}

\vbox{
\includegraphics[width=8.5cm]{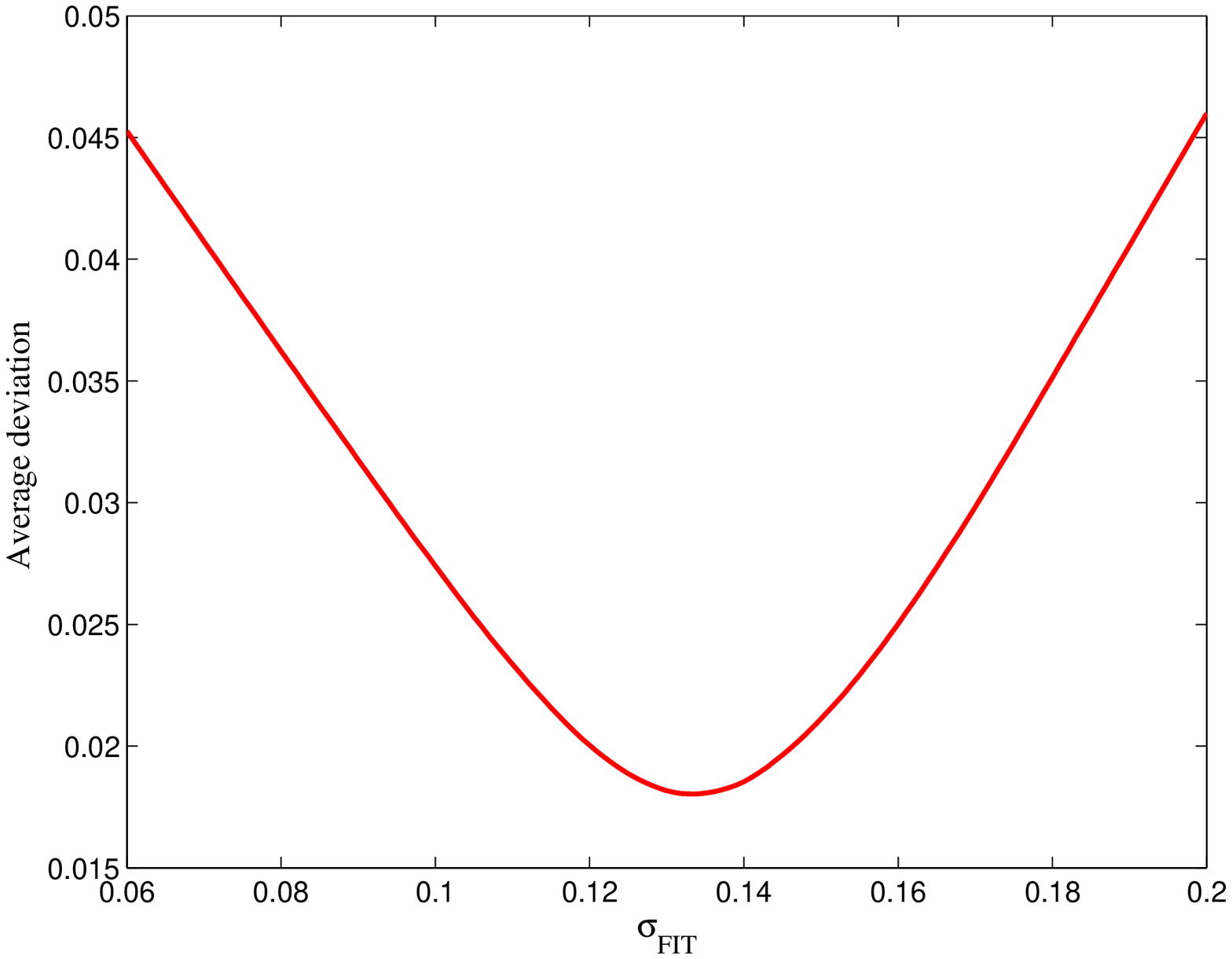}\\[1mm]
Average deviation of expected $\dH$ curves (according to
Eq.~\ref{GaussfitdH}) from $\dH$ dataset as a function of
$\sigma$. The best fit is obtained for $\sigma_{FIT} =0.133$,
which defines $\sest$.}

\vspace{0.4cm}

\vbox{
\includegraphics[width=8cm]{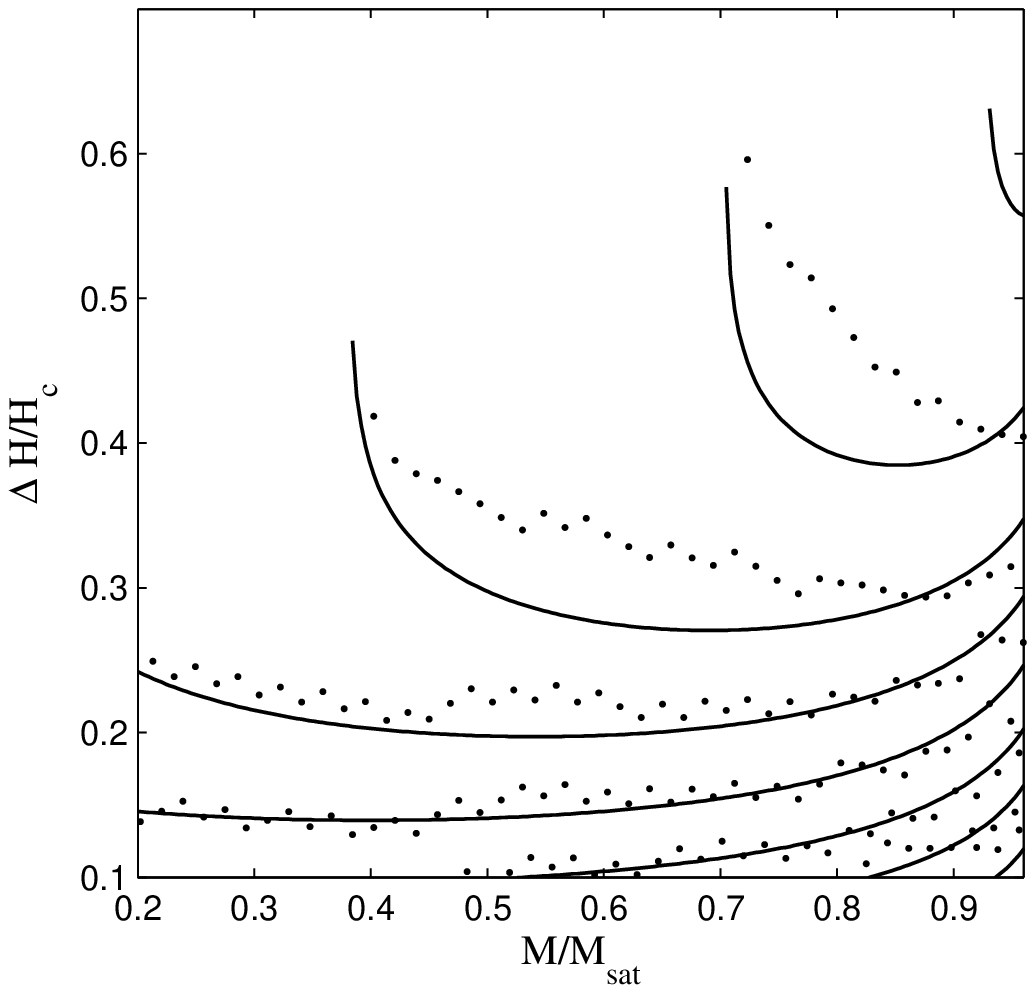}\\[1mm]
Comparison between expected $\dH$ curves (solid lines) for
$\sest=0.133$ and $\dH$ dataset (dots). Not all $\dH$ curves are
shown to keep the picture tidy.}

\end{document}